# To Send or Not to Send: An Optimal Stopping Approach to Network Coding in Multi-hop Wireless Networks

Nastooh Taheri Javan, Masoud Sabaei and Mehdi Dehghan
*{nastooh, sabaei, dehghan}@aut.ac.ir*

Computer Engineering and Information Technology Department
Amirkabir University of Technology (Tehran Polytechnic)
Tehran, Iran

**Abstract**

Network coding is all about combining a variety of packets and forwarding as much packets as possible in each transmission operation. The network coding technique improves the throughput efficiency of multi-hop wireless networks by taking advantage of the broadcast nature of wireless channels. However, there are some scenarios where the coding cannot be exploited due to the stochastic nature of the packet arrival process in the network. In these cases, the coding node faces two critical choices: forwarding the packet towards the destination without coding, thereby sacrificing the advantage of network coding, or, waiting for a while until a coding opportunity arises for the packets. Current research works have addressed this challenge for the case of a simple and restricted scheme called *reverse carpooling* where it is assumed that two flows with opposite directions arrive at the coding node. In this paper the issue is explored in a general sense based on the COPE architecture requiring no assumption about flows in multi-hop wireless networks. In particular, we address this sequential decision making problem by using the solid framework of optimal stopping theory, and derive the optimal stopping rule for the coding node to choose the optimal action to take, i.e. to wait for more coding opportunity or to stop immediately (and send packet). Our simulation results validate the effectiveness of the derived optimal stopping rule and show that the proposed scheme outperforms existing methods in terms of network throughput and energy consumption.
**Key words**: Multi-hop wireless networks; network coding; optimal stopping theory; coding opportunity.

## 1- Introduction

As a research topic, network coding theory is still in its early stages of development. Network coding was originally proposed by professor Ahlswede [1] that aimed to improve the resource efficiency of multicast communication in wired networks. The basic idea behind network coding is extraordinarily simple: instead of simple data forwarding, the intermediate nodes can combine the received packets before transmission. Among key properties of network coding is that it can reduce complexity and improve robustness and security, but the most desirable benefit of this technique is that it achieves significantly higher throughput. We can categorize the network coding operation into two classes: *intra-flow* network coding and *inter-flow* network coding [2]. In the first class, the packets that are coded together are from the same flow, but in the second class the coding operation combines packets from different flows.

In recent years a lot of studies have developed coding techniques to improve the capacity of wireless medium [3]. The majority of wireless links (except directional antenna) are broadcast links, and network coding can also be integrated with wireless broadcast to augment the information content per transmission. In this scenario, each wireless intermediate node can encode some of the received packets into one coded packet and then broadcast the coded packet to its neighbors through a single transmission. Among several proposed schemes, the study of Katti et al. [4] is a popular framework, called COPE, which allows coding between different sessions in multi-hop wireless networks. Although, the practical gain of network coding is lower than the theoretical gain [5], network coding application in wireless networks is a hot topic [6, 7].

The basic idea of network coding in wireless networks can be illustrated in Figure 1 [8], where node *N1* tries to deliver packet *P1* to node *N2* and *N2* tries to deliver packet *P2* to *N1*. Because of the limited radio range, they cannot communicate directly and they rely on an intermediate relay node to exchange packets. Without network coding *N1* sends *P1* to the relay, which forwards it to *N2*, and *N2* sends *P2* to the relay, which forwards it to *N1*. As a result, the total number of transmissions that are required for *N1* and *N2* to exchange their packets is four. But, in network coding structure, *N1* and *N2* could transmit their respective packets to the relay and relay XORs two packets together and broadcasts $P1 \oplus P2$. Upon receiving the coded packet, *N1* can decode *P2* by $P2 = P1 \oplus (P1 \oplus P2)$ and *N2* can decode *P1* by $P1 = P2 \oplus (P1 \oplus P2)$. Clearly, in this case only three transmissions are required for *N1* and *N2* to exchange their packets.

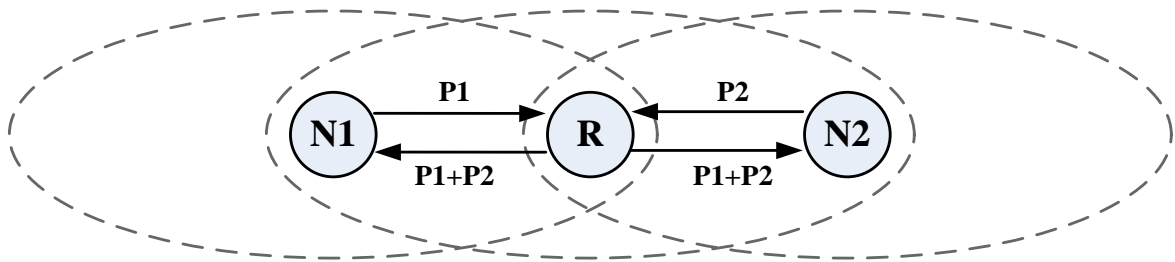

**Figure 1: Network coding in wireless networks [8].**

One of the challenges of network coding in a wireless environments is when the coding nodes has already some of the packets in its output queue, and that at this instance the node obtains the transmission chance from the MAC layer. In this case, if coding node cannot find a suitable coding option for its packets, it should decide whether to send the packet in head of line as a native packet right away or wait in hope of providing better coding options. Waiting for better coding opportunities introduces additional delay in packet transmission which is not desirable in many delay-sensitive applications. On the other hand, the coder node can wait for more favorable coding options in order to reduce the number of transmissions and consequently decrease power consumption. We called this dilemma the *to-send-or-not-to-send* in this paper. This problem as a sequential decision making problem, requires a trade-off between two performance criteria: end-to-end delay and the total number of transmissions. In fact, while decision to wait for more coding opportunities can decrease the number of transmissions, it is at the cost of adding to total network delay. Therefore, this decision directly affects the network performance and the benefits that can be achieved by network coding.

A glance at the background reveals that this topic has already been looked at [9-14] in a simple and restricted scenario called *reverse carpooling* [15]. Reverse carpooling, or *two-way* *relay* scenario, considers two flows which traverse a path in opposite directions in relay node. The reverse carpooling scenario is illustrated in Figure 2 for a simple relay network. The relay node *R* maintains two queues, $q_1$ and $q_2$ to store packets that should be delivered to the nodes $n_1$ and $n_2$, respectively. If none of the queues is empty, then *R* can create a coded packet by mixing the head-of-line packets from both queues. However, what should the relay node do if there is a packet in one of the queues, while another queue is empty? Let's now check this with latest available literature. For example [9] analysed this performance trade-off in this scenario under a simple FIFO strategy and it was concluded that the delay will tend toward ∞ if the energy saving was considered as the goal; while [10] put forward policies that make un-coded transmissions using certain probabilities; [11 and 12] approached it using waiting-time based policies, and finally [13] proposed an MDP solution for this trade-off in a restricted and very simple case with one packet per time-slot as transmission capacity at the relay node.

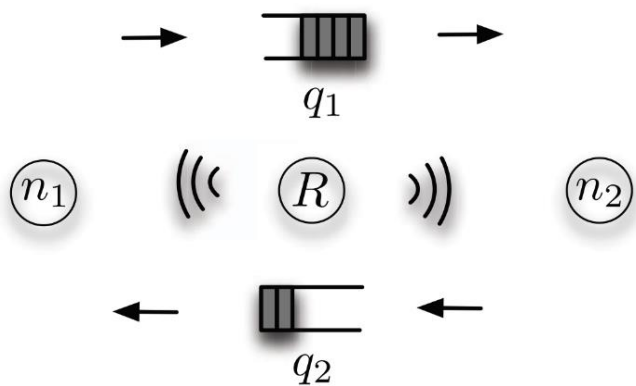

**Figure 2: Reverse carpooling scenario [14]**

The main point in all of the aforementioned studies is that authors boiled down the problem to reverse carpooling (two-way relay) scenario that is not a general scheme for multi-hop wireless networks. For example, we know that in reverse carpooling, the coding option is predesignated and doesn't change with time. This snag can be resolved, as demonstrated in our presentation, by taking a general approach based on a well-known coding architecture called COPE [4] which is the first practical network coding based packet forwarding architecture to improve network throughput of wireless multi-hop networks.

In COPE, each network node overhears its neighbors' traffics opportunistically, saves the overheard packets in its memory and then announces the list of packets in its memory to its neighbors. Announcing this overhearing information can be done using two methods: one is by explicitly acknowledging the overhearing information, and the other is by obtaining overhearing information statically [4]. In the first approach, a network node sends a list of all the packets in its memory to its neighbors periodically as reception reports, whereas in the second approach, the overhearing information is guessed by the encoding node through using the link quality advertisements via periodic probing. Comparisons of different marking approaches and their performances are beyond the scope of this paper. We opted for the reception report approach in our proposed scheme, which are piggy-backed over normal data packets. In particular, in this method special packet headers should be added to the data packet header between the network and the data link headers, which is sometimes known as *coding header*. In this conditions, the coding node can intelligently find the best coding option (with highest *coding degree*) and encode multiple packets destined to different next hops. The number of packets that can be encoded by a coding node in each transmission is defined as the coding degree of that coding option.

In this paper, we formulate the to-send-or-not-to-send problem as an optimal stopping problem [16] and derive an optimal stopping rule for coding nodes. In our scheme, no special assumption is considered for available flows (either in terms of the number of flows or the flows' direction) and any node separately reviews its coding opportunities and selects the best coding option. Specifically, in this model, each node checks its coding opportunities for the available packets in its output queue based both on the information about the packets in its neighbors' buffer pools and the packets to be received by its neighbors, and then discovers the best coding option. In each transmission opportunity, a coding node can decide whether to encode and transmit the coded packet by applying selected coding option, or to wait until the next transmission slot in order to get better coding opportunities. In particular, before while waiting for a transmission opportunity to occur and in coding node's idle periods, the coding node checks the available coding options for the first packet in sending queue and selects the best option for it. Therefore, when a transmission opportunity occurs, the node knows both the best coding option and the best coding degree for the ready-to-send packet and uses this value in the decision making process. If the node has decided to wait, received data packets and the reception reports from the neighbors may provide better coding options for packets in the output queue with higher coding degree values. Our objective then is to decide *when* the coding node should stop waiting and transmit the coded (or native) packet.

As mentioned before, there is a fundamental trade-off between number of transmissions and data packets end-to-end delay. The desired trade-off is essentially formulated as an optimal stopping problem and boils down to choosing the optimal stopping rule for transmission time, in order to reduce the total number of transmissions. In fact, based on the observations collected up to a decision epoch, i.e., transmission opportunity, the optimal stopping problem seeks to find the *best* time to stop observing and transmit a packet. In our proposed formulation, using 1-SLA (one-stage look-ahead) approach [17], finding the optimal policy is reduced to simply finding a *threshold* on the coding degree of packets before initiating a transmission. The optimality of the 1–SLA rule is analytically validated by Ferguson [17]; who demonstrated that in a monotone optimal stopping problem, the 1-SLA rule is optimal. From the definition, the 1-SLA rule is the one that stops if the reward for stopping at the current step is at least as the same as the expected reward of continuing only one step before stopping. We will prove that our scheme satisfies the 1-SLA conditions and explain how to derive a stopping rule for transmission in general case for wireless multi-hop networks.

The rest of this paper is proceed as follows. In Section 2, we examine the literature on the to-send-or-not-to-send like problems solutions that tap reverse carpooling scheme to find a solution. In Section 3, we describe our system model in detail from the network and coding viewpoints. In Section 4, using the optimal stopping theory, we formulate the problem and derive a stopping rule for coding nodes. Section 5 gives the numerical results and Section 6 concludes this paper and recaps its contribution.

## 2- Related Works

In recent years, some researches have been made in the study of delayed transmission in network coded wireless systems. A survey of related literature reveals that these attempts have mostly dealt with the challenge similar to the to-send-or-not-to-send problem and that they fall into two main categories: most of them considered this problem in reverse carpooling scenario and picked out the delayed transmission solution. In contrast, few have tried to solve the problem with an approach other than delayed transmission (such as change congestion control mechanism!)

In [14] an on-line approach was proposed to make transmit/wait decisions at the intermediate nodes in reverse carpooling scheme with a single relay node. This algorithm minimized the total system cost in terms of both the delay and number of transmissions based on the primal-dual method. Authors in [18], modeled the two-way relay coded wireless networks as a continuous time Markov chain and showed that two-way relay networks that employ network coding are associated with negative and positive customers problem. They analyzed the energy consumption of wireless networks under this scenario and then they provided upper and lower bounds on the energy consumption for this situation. In [13] the technique came up with an energy-delay trade-off model based on MDP formulation in a reverse carpooling scenario with a maximum transmission capacity of one packet per slot at the relay node. In [19], focus also was put on costs for transmission and delay in reverse carpooling scenario and authors also formulated the problem as an MDP. They argued that the optimal policy is threshold type and obtained it through modelling the system as a Markov chain. In contrast, their counterparts in [12] analyzed the energy-delay performance using waiting-time based policies in two-way relay scheme. Authors in [10] formulated delay and energy consumption by considering Markov Chain and a Hidden Markov Model in their proposal. They also developed a policy with minimal average delay and zero packet-loss rate. In this case, each time slot was split into two parts, where the packet to be relayed were received in the first part, and sent in the second part. They characterized the buffer state by using a finite state Markov chain, which allowed them to investigate the network layer performance, namely, delay, packet loss and power consumption. In the aforementioned studies, it has been invariably assumed that all packets, whether coded or non-coded, are received with 100% accuracy. However, in [20], authors took the view that quality of channel would change over time and developed a practical solution based on two-way relay fading channels. They used MDP framework that considered both the wireless channel state and the transmission queue.

In some other researches authors considered more restricted conditions and scenarios. For example, in [11] researchers explored a scheme in which two users could communicate in network ceded environment through a single powerful access point. They developed a theoretical framework using a Discrete Time Markov Chain (DTMC) to model the system and Discrete time Markovian Arrival Process (DMAP) to model packet arrivals. Finally, they found the age distribution of the waiting packets and hence determined the waiting-time which achieved the optimal trades-off between spectrum access efficiency and packet delay. Authors in [21] assumed that a global knowledge of queue backlogs was available and developed game-based distributed strategies for optimizing the delay-energy performance in a two-way relay coded network. They then determined the Nash equilibrium of the game and discovered that it performed worse than the centralized algorithm. To overcome the challenge, they introduced a pricing strategy at the relay. In [22], Mohammad and Farid assumed that each node was assigned a frequency band orthogonal to the others (FDMA). In order that each node could send and receive data concurrently, the frequency division duplexing (FDD) was configured for data transmission through each source-relay link. Finally, they proposed three network coding schemes based on power-delay constraint also in two-way relay networks. Sgduyu and Ephermides in [23] restricted their analytical studies to a simple tandem line network and designed the cross-layer scheme between MAC layer and network layer and incorporated some trade-offs in their proposal. In [24], authors proposed a video-aware packet delaying transmission mechanism for video streaming application in relay nodes in order to overcome video traffic bursty nature. In their solution the source attaches a tag on each packet to indicate the QOS level for intermediate nodes which can be used to make decision about the packets.

Finally, it is also worth mentioning that a number of researches attempts to tackle this issue, some studies have put forward an entirely different solution compared with delayed transmission. For example, in [25], authors concluded rate mismatch between TCP flows, due to the dynamic nature of TCP, can decrease the coding opportunities dramatically, as there might not be enough packets from different sessions at relay nodes to encode together. Based on this viewpoint, authors in [26] proposed a modified congestion control mechanism in relay nodes. In their queue management algorithm, the packets was dropped from the queue based on both network coding and congestion state. In Table 1, we summarize the general aspects of related works that have investigated a problem similar to the to-send-or-not-to-send. As mentioned before, all of the referred works boiled down this problem to restricted scenarios (such as reverse carpooling), whereas we study a more general form of this problem with no special assumptions about flows in multi-hop wireless networks.

**Table 1. Comparison of some related works in the literature**

| Ref. # | Main Configuration and Assumptions | Main Objective | Theoretical Framework | General Solution |
|---|---|---|---|---|
| 14 | reverse carpooling | trade-off between no. of transmissions & delay | Primal dual optimization | delayed transmission in relay node |
| 18 | reverse carpooling | finding bounds for energy consumption | Continuous Time Markov Chain | - (just analytical work) |
| 13 | reverse carpooling | trade-off between no. of transmissions & delay | MDP | delayed transmission in relay node |
| 19 | reverse carpooling | trade-off between no. of transmissions & delay | MDP | delayed transmission in relay node |
| 12 | reverse carpooling | trade-off between energy consumption & delay | Constrained linier optimization | - (just analytical work) |
| 10 | reverse carpooling | trade-off between energy consumption & delay | HMM | delayed transmission in relay node |
| 20 | reverse carpooling with fading channel | trade-off between coding opportunity & delay | MDP | delayed transmission in relay node |
| 11 | reverse carpooling with stronger relay | trade-off between channel efficiency & delay | DTMC | delayed transmission in relay node |
| 21 | reverse carpooling with global knowledge | trade-off between cost & delay | Game Theory | delayed transmission in source nodes |
| 22 | FDMA based reverse carpooling | trade-off between energy consumption & delay | Markov Chain | new coding structure |
| 23 | reverse carpooling tandem networks | minimizing energy costs in tandem networks | Linier optimization | design a cross layer mechanism |
| 24 | video streaming application | trade-off between coding opportunity & delay | Linier optimization | delayed transmission in relay node |
| 26 | TCP traffic applications | trade-off between no. of transmissions & delay | NUM | new congestion control method |

# 3- System Model and Assumptions

In this section we described our system model and assumptions, from the network model and coding model perspectives. Some of key assumptions are numbered for easier references.

## 3-1- Network Model

We consider a stationary multi-hop wireless network [27], supporting multiple unicast sessions. We assume in our proposal that each node can send and receive data via a half-duplex scheme and also that nodes share the common channel bandwidth with each other. In these assumptions, all network nodes are equipped with omni-directional antennas and support the same maximum transmission rate. Network nodes are distributed in a two-dimensional region randomly and can communicate with their neighbors placed in a radius $\rho$. Each node in the network can be either a source of or a destination of a flow and all packets have the same size.

We assume the network nodes use the IEEE 802.11 as MAC layer protocol which is the main standard based on the carrier sense multiple access scheme with collision avoidance (CSMA/CA). The protocol requires a node to sense the medium before attempting transmissions. A node with a new packet to transmit monitors the channel activity. If it is idle, the node can start transmission, and we say the node achieves one *transmission opportunity*. A sender node selects a neighbor as destination and sends a packet to it via unicast by including its address in the packet header. All other neighbors are in promiscuous mode and can thus overhear the transmission. As defined in CSMA/CA, the nodes use RTS/CTS handshake.

In practice, the encoded packets require all the next-hops to acknowledge the receipt of the associated native packets. Since we use local retransmissions, so, the sender expects the next-hops of an encoded packet to decode the coded packet, obtain their native packet, and ACK it. If any of the native packets is not ACKed within a certain time interval, the packet is retransmitted, potentially encoded with another set of native packets. Unfortunately, extending the pure synchronous ACK approach to coded packets is highly inefficient, as the overhead incurred by sending each ACK in its own packet would be excessive. We assumed (as does COPE) that the encoded packets are ACKed asynchronously. In particular, when a node sends an encoded packet, it schedules a retransmission event for each of the native packets in the encoded packet. If any of these packets is not ACKed within a specific period, it is inserted at the head of the output queue and retransmitted. Retransmitted packets may get encoded by other packets. A next-hop that receives an encoded packet decodes it to obtain its native packet, and immediately schedules an ACK event. Before transmitting a packet, the node checks its pending ACK events and incorporates the pending ACKs in the header.

**Assumption 3-1:** We assume that *T*, the duration between two consecutive transmission opportunities for a particular node, has an exponential distribution with the parameter $\lambda_t$.

For a given wireless node, the transmission opportunities are abundantly available whenever there is less or none transmission to the other nodes. This scenario is rarely possible because the channel is a shared medium, and usually everyone wants to transmit. In practice, the availability of transmission opportunities for a node can be dependent on multiple factors: network topology, local congestion, neighbors' traffic pattern, etc. So, the transmission opportunities arrival times can be considered as random variables. To understand distribution of the transmission opportunities arrival times, we present our simulation results and show the recorded inter-arrival times of transmission opportunities of nodes. The simulation environment consists of a multi-hop wireless network employing 200 stationary nodes, (our simulation environment was described in detail in Section 5-1). In this scenario we only focus on MAC layer transmission opportunities, and generate some traffic randomly across the whole network. By evaluating the results, we found that the distribution of inter-arrival time between two successive transmission opportunities fits exponential distribution. Figure 3 shows Quantile-Quantile plots [28] for the distribution of time between transmission epochs and an exponential distribution with parameter lambda=58.19. From these results we conclude that the transmission epochs inter-arrival times closely resemble an exponential distribution.

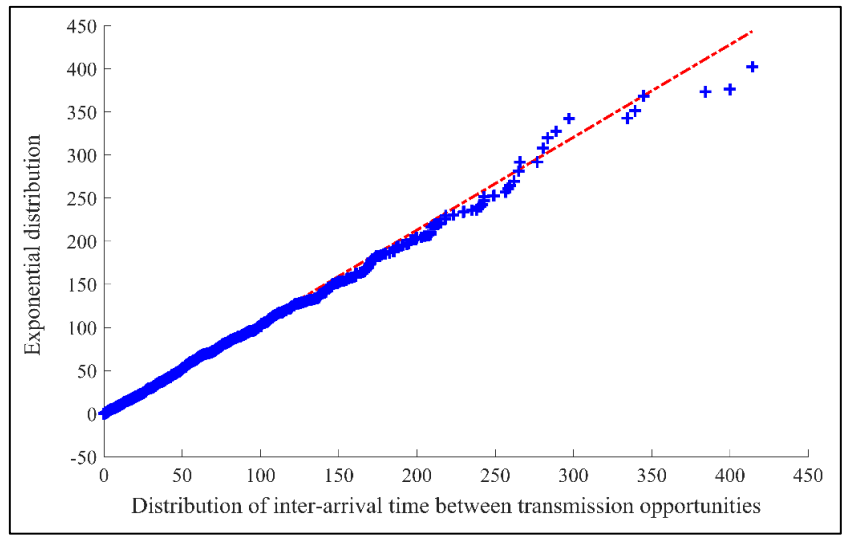

**Figure 3: The QQ-plot of the transmission opportunities distribution**

**Assumption 3-2:** We assume that packets arrive at a node according to a homogeneous Poisson process with intensity $\lambda_p$ [29].

In this study, we assumed a traffic model could be defined as follows: Each node in the network could be a source, destination and/or relay of packets. Also, for the sake of simplicity, we further assumed packet generation process at each node was an i.i.d. Poisson process. It should be noted, however, that our analysis could be easily extended to any arbitrary packet arrival process since the diffusion approximation method applies to non-Poisson arrival process as well, but other expressions would become more complex. Finally, at each node, packet arrivals are Poisson with rate $\lambda_p$ (in packets per second). It should be noted that the number of received packets during a specific time horizon varies from zero to infinity. Also, the rate at which a node receives packets is constant with respect to time and therefore, we might as well accept that the number of received packets by a specific node during a time horizon $t$, follows a Poisson distribution with parameter $\lambda_p.t$.

## 3-2- Coding Model

We further assume that coder nodes use the COPE framework [4] for one-hop opportunistic network coding. In this model, each node *i* overhears all packets in its neighborhood, places the received packets in its memory, and announces the list of its packets to its neighbors by broadcasting reception reports. Each node has two separated queues: transmission queue and overheard queue. The coder node will code the ready-to-transmit packet along with other packets either in its transmission queue or overheard queue based on the coding conditions by XORing the packets together. In this case, when a node receives a coded packet, it retrieves the corresponding packet with the help of other packets in its memory. To do this, the node XORs the (*n*-1) packets with the received encoded packet to retrieve the native packet intended for it.

Whenever a transmission opportunity arises and the coding node decides to stop (i.e., transmit), the packet with highest coding degree must be transmitted, either in an encoded way or in a native way (by coding degree 1). Additionally, we suppose that the required overhead of transmitting coding coefficients is negligible.

**Assumption 3-3:** We assumed that reception reports are received based on a Poisson distribution with parameter $\lambda_r$. This assumption is based on Assumption 3-2. Since each data packet can carry a reception report, we can assume reception reports arrivals have Poisson distribution.

Actually, coding degree for packets in output queue increases due to the effect of following two factors: (i) arrival of new packets to the node's output queue, which allows new coding opportunity to be created for sending packets, (ii) reception reports from neighbors, which results in new coding opportunity. In particular, each received data packet, irrespective of subsequent packets, increases coding degree of the best coding option by 1 with probability $p_p$ and each received reception report, independently increases *d*, the best coding option degree, by 1 with probability $p_r$. Given Assumption 3-2, the amount of increase in coding degree of the best coding option which is as a result of receiving new packets follows a Poisson distribution with parameter $\lambda_p.t.p_p$. Therefore, $\lambda_d$, the overall rate of increase in *d*, can be calculated as follows:

$$\lambda_d = \lambda_r.p_r + \lambda_p.p_p \qquad (1)$$

Generally, it is very difficult to specify a closed form for parameters $p_p$ and $p_r$ and model them in multi-hop wireless networks. This is because, values of $p_p$ and $p_r$ depend on a lot of various parameters such as network topology, number of neighbors, traffic pattern in coding node and its neighbors, number of flows, buffer maps in coding node and its neighbors, etc. To the best of our knowledge, to date, no parameters similar to $p_p$ and $p_r$ have been modeled or formulated in closed form. Finally, whenever $p_p$ and $p_r$ will be formulated in the closed form model, they can be inserted in equation (1). So, in practice, the node should estimate $\lambda_d$ online from local observation by using simple *adaptive filters* [30] such as LMS [31].

## 4- Optimal Stopping Strategy

In this part, we describe our modelling strategy by optimal stopping approach. In Section 4-1, we present the problem formulation and in two successive sections we describe how to derive a stopping rule.

### 4-1- Problem Formulation

In recent years, researchers facing sequential decision-making problems in which an agent has to select the optimal time to perform an action based on a chain of observed random variables, have come up with ways to formulate an optimal stopping theory [32]. In particular, optimal stopping problems can be defined via two objects:
- a sequence of random variables: {$X_1, X_2,...$} with known joint distribution;
- a sequence of real-valued reward functions: {$y_0, y_1(x_1), y_2(x_1, x_2),...,y_\infty(x_1, x_2,...)$}.

Given the aforementioned objects, the optimal stopping problem can be explained as follows [17]. The decision maker observes the sequence of random variables $X_n = x_n$, and at each decision epoch, the agent can select to either stop observing and get the known reward $y_n(x_1,...,x_n)$ or continue and observe the next variable $X_{n+1}$. In fact, optimal stopping problems are to select the best time to stop in order to maximize the expected reward (or minimize the expected cost), which means that a decision maker has to control the stopping time $0 \leq N \leq \infty$ that maximizes $E[Y_N]$, where $Y_N = y_N(x_1,...,x_N)$ indicates the random reward associated with stopping at $N$, and $E[.]$ is related to the expected value function. The proposed strategy formulation is going to be based on a decision to transmit the encoded (or even native) packets right away or wait for better coding options to come along, hence the problem can be modeled as an optimal stopping problem.

In our case, the sequential decision making problem associated with a node is to decide the next action (continue to wait for a better coding option or stop immediately) at each decision point, and a decision horizon starts at the beginning of the next packet transmission operation. When a decision for stopping has been made, the current packet is sent out (in form of native or coded) and the next decision horizon starts for the next packet transmission. In Table 2, the correspondence between the optimal stopping theory and the proposed problem is illustrated.

**Table 2: The correspondence between the optimal stopping theory and our problem**

| | | |
|---:|:---:|:---|
| decision maker | ⟷ | coding node |
| observation | ⟷ | coding opportunities |
| variables value function | ⟷ | encoding degree |
| loss function | ⟷ | total packets delay in a coding node |
| stop | ⟷ | transmit |
| decision epoch | ⟷ | transmission opportunity |

We can model our problem with a 4-tuple $<S, A, R, P>$ where $S$ is state space, $A$ is action set, $R$ is reward function and $P$ is state transition probability matrix.

**Stage:** A stage is the duration between two consecutive transmission epochs and the decision point is exactly the moment when the opportunity of transmission presents itself. So, at each available transmission opportunity, the node decides to either wait for a better coding option or send (stop).

**State variable:** The state of the node, $d$, according to which the decision must be based, is the *best* possible coding option available for all packets in output queue and $D = \{1, 2, 3, ...\}$ presents all the possible states. $d$ is dependent of the current node's output queue, other nodes' packet pools, and next-hops of packets in the current node's queue, so all these factors must be taken into account when determining $d$'s distribution function.

**Action Set:** At each decision epoch, the node decides to either 0) wait for a better coding option or 1) send (stop), so $A$, action set, is $\{0, 1\}$.

As has already been discussed, the challenge in all optimal stopping problems, is one of deciding to stop or to continue an action. Here, stopping means to transmit the encoded (or native) packets after which the system is immediately transferred to state 1; in other words, after transmitting a packet, $d$ is immediately set to 1 for the next decision epoch. If we decide to stop and transmit the packets, we are going to face a new optimal stopping problem in the next epoch. Now a decision to continue, means the node is able to receive more packets or reception reports, and therefore $d$ might be improved.

**Transition Matrix:** Let $p_{ij}(a, T_0)$ denote the transition probability from state *i* to state *j* when action *a* is chosen and the next stage happens in $T_0$ units of time, then the transition matrix can be defined as follows:

$$P = [p_{ij}(a, T_0)] \qquad (2)$$

Now by taking *D* to represent the set of all possible states we have:

$$p_{ij}(1, T_0) = p\{x_{t+1} = j | x_t = i, a = 1, T = T_0\}; i, j \epsilon D \qquad (3)$$

$$p_{ij}(1, T_0) = 0; \quad \forall j \epsilon D - \{1\} \qquad (4)$$

$$p_{i1}(1, T_0) = 1; \qquad \forall i \epsilon D \qquad (5)$$

The above equations imply that choosing action 1 (transmission of the encoded packets) transits the system to state 1 (where *d* = 1) and leaves the system in that state. State 1 is in fact an absorbing state. Also we have:

$$p_{ij}(0, T_0) = p\{x_{t+1} = j | x_t = i, a = 0, T = T_0\} = \begin{cases} 0 & ; j < i \\ \dfrac{e^{-\lambda_d T_0}(\lambda_d T_0)^{j-i}}{(j-i)!} & ; j \geq i \end{cases} \qquad (6)$$

which implies that choosing action 0 (waiting for another epoch) in each pulls the system to states more desirable condition (one with higher *d*), where continuing to wait and receive packets does not decrease *d*.

**Reward function:** Let $R(d, a)$ be the amount of reward achieved by selecting action *a* while in state, *d*, which is defined as follows:

$$R(d, a) = \begin{cases} g(d-1) & ; a = 1 \\ 0 & ; a = 0 \end{cases} \qquad (7)$$

here, $g(.)$ is the gain achieved by the reduced number of transmissions. In wireless networks when a coded packet is transmitted with coding degree *d*, it is exactly reduced *d*-1 transmissions in the network, so $g(.)$ is non-negative and non-decreasing with respect to *d* and assumed to be finite. In other words:

$$E\{g(d-1)\} < \infty \qquad (8)$$

In case of postponing the transmission to the next decision epoch, the system incurs a penalty cost associated with the encoded packets delay. This penalty cost is not accumulative and as soon as the packets are transmitted, the gain will be decreased by a factor due to the penalty costs of the packets. This fraction is assumed to be $e^{-L.\delta.T_n}$ in which $T_n$ represents the total waiting time of the encoded packets, *L* indicates the maximum allowable number of output queue (buffer size) and $\delta$ is a discount factor. Therefore, the random gain sequence can be formulated as follows:

$$Y_0 = 0$$
$$Y_n(d_1, d_2, \dots, d_n, T_n) = g(d_n - 1) \times e^{-L.\delta.T_n}; n \geq 1 \qquad (9)$$

In (9), the achieved coding gain is discounted by the delay experimented in coding. We use the exponential discount of the reward because exponential discount is monotone and can monotonically decrease the reward. This structure is helpful for developing control policies.

Moreover, the exponential discount factor can also handle the additive delay and the commonly used reward *g*(*state*)-*T* in literature can be easily translated into the exponential structure [33].

**Assumption 4-1:** Let's assume we decide to continue, then for each $d \epsilon D$ in the $n^{th}$ transmission epoch the time interval before the next transmission opportunity and the amount of increase in *d* are both independent of the previous transmission epochs.

Using Assumption 4-1, this problem can be formulated as a Markov Decision Process (MDP) and based on the decision set it can be modeled as an optimal stopping problem. Our formulation notation is summarized in Table 3.

**Table 3: Summary of Notations**

| | |
|---|---|
| $d$ | state of a node (current best coding option degree) |
| $d^*$ | the optimal coding degree for stop |
| $\lambda$ | arrival rate of data packet |
| $\lambda_r$ | arrival rate of reception reports |
| $\lambda_d$ | average increment of best coding option degree per time unit |
| $\lambda_t$ | average number of transmission opportunity per time unit |
| $p$ | best coding option degree increment probability by each received data packet |
| $p_f$ | best coding option degree increment probability by each received reception report |
| $T$ | duration between two consecutive transmission opportunities |
| $\delta$ | delay discount factor for exponential reward function |
| $L$ | the node buffer size |
| $p_{i,j}$ | transition probability from state *i* to state *j* |

## 4-2- Optimal Stopping Rule

In our problem, finding the optimal policy is reduced to simply finding a threshold on the coding degree of packets by 1-SLA approach. Ferguson illustrated in [17] that in monotone optimal stopping problems, the 1-SLA rule is optimal.

**Theorem 1.** *In optimal stopping problems, if $E\{\sup_n Y_n\} < \infty$, $\limsup_{n \to \infty} Y_n \leq Y_\infty$ and the problem is monotone, then the 1-SLA rule is optimal.*

*Proof:* See Chapter 5 of [17]. ∎

We will now proceed to the next step of proving the proposed scheme satisfies the 1-SLA conditions in Theorem 1 and then derive a stopping rule for delayed transmission in general case for multi-hop wireless networks.

***Lemma 1.*** In our problem, two conditions $E\{\sup_n Y_n\} < \infty$ and $\limsup_{n \to \infty} Y_n \leq Y_\infty$ are satisfied.

*Proof*: In the proposed scheme, the first condition does not hold except when the time interval between $n-1^{th}$ decision epoch and the $n^{th}$ decision epoch is infinite. In this case, the increase both in *d* and gain, due to reduced number of transmission, will be infinite. Since it is assumed that this time interval follows an exponential distribution, the above probability would always be equal to 0 and, consequently, the duration of intervals between decision epochs would be finite. The second condition states that if the process continues to infinite stages (i.e. over infinite number of stages we do not decide to stop), we will end up with a gain that is the finite amount of $Y_\infty$. In our proposed solution, if the packets do not get transmitted over infinite stages, the overall penalty cost would will be very high and hence, transmitting after this period achieves no gain at all, i.e., $Y_\infty = 0$. These statements can be formulated as follows:

$$\limsup_{n\to\infty} Y_n = \lim_{n\to\infty} g(d_n - 1) \times e^{-L.\delta.T_n} \tag{10}$$

where:

$$\left.\begin{array}{l}\lim_{n\to\infty} e^{-L.\delta.T_n} = 0 \\ \lim_{n\to\infty} g(d_n - 1) < \infty\end{array}\right\} \Rightarrow \lim_{n\to\infty} g(d_n - 1) \times e^{-L.\delta.T_n} = 0 \tag{11}$$

∎

In optimal stopping problems, the optimality equation can be defined as follows [34]:

$$v(i) = \max\left\{R(i), -C(i) + \sum_j p_{ij} v(j)\right\} \tag{12}$$

where *v(i)* is the minimum value for optimality equation if the system is in state *i*. *R(i)* is the reward while the system stops in state *i*, *C(i)* indicates the amount of cost for transition to next state *j*, and *p<sub>ij</sub>* denotes transition probability from state *i* to state *j*. Now, with reference to all earlier assumptions and system model, the optimality equation for our solution can be written as:

$$v(d) = \max\left\{0 + \int_0^\infty \sum_{j \geq d} p_{dj}(0,T) \times v(j) \times e^{-L.\delta.T} f_T(t)dt, g(d-1) \right.$$
$$\left. + \int_0^\infty p_{d1}(1,T) \times v(1) f_T(t)dt\right\} \tag{13}$$

where *v(d)* is the maximum expected gain while the system's state is *d*. In the above optimality equation, the first term captures the expected achievable gain if we continue to the next stage, and the second term indicates the what that will be as a result of transmitting the encoded packets. When *d* = 1, deciding to transmit the packets does not reduce the number of transmissions so:

$$v(1) = 0$$

By simplifying the equation (13) we have:

$$v(d) = \max\left\{\int_0^\infty \sum_{j \geq d} p_{dj}(0,T) \times v(j) \times e^{-L.\delta.T} f_T(t)dt, g(d-1)\right\} \tag{14}$$

### 4-3-Finding the Optimal Solution

To obtain an optimal stopping solution, we have to work out compute not only for how long it is beneficial to wait but also under what condition it is no longer reasonable to wait. In other words, we are stating that it is optimal to stop at a particular stage if the reward is no less than the expected reward of stopping at a subsequent stage. The way to figure this out is to compare gain for stopping in current state versus the expected reward for waiting. So, we define *B* set as follows:

$$B = \left\{ d : g(d-1) \geq \int_0^\infty \sum_{j \geq d} p_{dj}(0,T) \times g(j-1) \times e^{-L.\delta.T} f_T(t) dt \right\} \quad (15)$$

where $B$ represents the set of all the states, for which, stopping in one stage is at least as desirable as continuing to the next stage and stopping and $g(d-1)$ is the reward by the coding node for stopping in current stage and $\int_0^\infty \sum_{j \geq d} p_{dj}(0,T) \times g(j-1) \times e^{-L.\delta.T} f_T(t) dt$ represent the expected reward if it waits until the next stage and then stops. By simplifying equation (15), we have:

$$B = \left\{ d : g(d-1) \geq \int_0^\infty \sum_{j \geq d} \frac{e^{-\lambda_d T}(\lambda_d T)^{j-d}}{(j-d)!} \times g(j-1) \times e^{-L.\delta.T} \times \lambda_t e^{-\lambda_t T} dT \right\} \quad (16)$$

Next, we move on to prove that $B$ is a closed set on $d$ and then use the 1-SLA approach to devise our stopping rule. In fact, when a coded packet is transmitted with coding degree $d$, it is reduced $d-1$ transmissions in the network, thus by assuming $g(.)$ to be in the linear form of $g(d) = cd + b$, the set $B$ would be as follows:

$$B = \left\{ d : cd + b \geq \int_0^\infty \sum_{j \geq d} \frac{e^{-\lambda_d T}(\lambda_d T)^{j-d}}{(j-d)!} \times (cj + b) \times e^{-L.\delta.T} \times \lambda_t e^{-\lambda_t T} dT \right\} \quad (17)$$

Simplifying the above statement results in the following relationship:

$$B = \{d : cd + b \geq cd.E_T(e^{-L.\delta.T}) + c\lambda_d E_T(Te^{-L.\delta.T}) + b.E_T(e^{-L.\delta.T})\}$$
$$= \{d : (1 - E_T(e^{-L.\delta.T}))(cd + b) \geq c\lambda_d E_T(Te^{-L.\delta.T})\}$$
$$= \left\{ d : d \geq \frac{\lambda_d E_T(Te^{-L.\delta.T})}{(1 - E_T(e^{-L.\delta.T}))} - \frac{b}{c} \right\} \quad (18)$$

Now by further simplifying $E_T(Te^{-L.\delta.T})$ and $E_T(e^{-L.\delta.T})$ we have:

$$E_T(e^{-L.\delta.T}) = \int_0^\infty e^{-L.\delta.T} \times \lambda_t e^{-\lambda_t T} dT = \frac{\lambda_t}{\delta L + \lambda_t} \quad (19)$$

and

$$E_T(Te^{-L.\delta.T}) = \int_0^\infty Te^{-L.\delta.T} \times \lambda_t e^{-\lambda_t T} dT = \frac{\lambda_t}{(\delta L + \lambda_t)^2} \quad (20)$$

Finally, by considering (19) and (20), the equation (18) is simplified and the set $B$ is determined as follows:

$$B = \left\{ d : d \geq \frac{\lambda_d \frac{\lambda_t}{(\delta L + \lambda_t)^2}}{\left(1 - \frac{\lambda_t}{\delta L + \lambda_t}\right)} - \frac{b}{c} \right\}$$

$$= \left\{d : d \geq \lambda_d \frac{\lambda_t}{\delta L(\delta L + \lambda_t)} - \frac{b}{c}\right\} \tag{21}$$

***Lemma 2.*** For our scheme, the stopping rule (21) is the optimal solution to Problem (14).

*Proof:* It can be simply proved that *B* in (21) is a closed set on *d* and is monotone since the right side of the above equation, $\lambda_d \frac{\lambda_t}{\delta L(\delta L + \lambda_t)} - \frac{b}{c}$, is a constant and the left side is not constant and can increase (*d* does not get worse in case of not transmitting), consequently, the above inequality always holds if we decide to wait. As long as *B* is a closed set, and by considering *Lemma 1*, the 1-SLA rule in equation (21) is the optimal stopping rule according to Theorem 1. ∎

So, the optimal solution to the problem would be:

$$d^* = \lambda_d \frac{\lambda_t}{\delta L(\delta L + \lambda_t)} - \frac{b}{c} \tag{22}$$

Finally, *w(d)*, the decision rule, would be:

$$\boldsymbol{w(d)} = \begin{cases} 0; & d < d^* \\ 1; & d \geq d^* \end{cases} \tag{23}$$

Intuitively, we can say $d^*$ is affected by two network parameters which are: transmission rate at the MAC layer and coding opportunities creation rate. If coding opportunities rate is high, the value that is obtained for $d^*$ will be large which means waiting in hope of better coding degree is reasonable. Now, let's assume there is little transmission opportunities at the MAC layer, since the time interval between two consecutive transmissions will be very long, this will result in low values for $d^*$ and the coding node has to be contend with a low value of coding degree; It is important to note that since packet cost in terms of delay is high, under this circumstances, it isn't logical to wait for more coding opportunities to arise. On the other hand, where transmission opportunities at the MAC layer is significant, we will end up with larges values for $d^*$ and, therefore, the coding node can afford to miss some opportunities in hope of better coding degrees.

**Remark 1**. As was mentioned before, we assume that an increase in the value of *d* is the result of receiving new data packets or receiving neighbors' reception reports and we cannot specify a closed form for the parameters $p_p$ and $p_r$ because of their dependencies on various other parameters. Knowledge of these parameters is thus required in order to evaluate equation (1). If this knowledge is not available a priori, $\lambda_d$ can only be accurately estimated by a sufficiently large number of measurements. Actually, even when $p_p$ and $p_r$ are known, it is rarely possible to obtain $\lambda_d$ in closed form. Fortunately there is an adaptive estimation algorithm based on LMS (Least Mean Square) [31] that can be utilized for $\lambda_d$ value estimation. Adaptive filters [30] are typically used in applications in which signals with unknown statistics are involved. In this case, the LMS filter is employed to predict an estimation $U_p[n]$ of $\lambda_d$ at the *n*+1 step, as a linear combination of $\{U[n-1], U[n-2], \ldots, U[n-k]\}$ by:

$$U_p[n+1] = \sum_{k=0}^{Z-1} h_n[k] * U[n-k] \tag{24}$$

where *U[n]* denotes the average normalized $\lambda_d$ in the interval between $n^{th}$ and $n$-$1^{th}$ observation periods and each *U[n]* value is weighted by the accordant filter coefficient $h_n[k]$. Using this technique, prediction error is calculated by $U_e[n] = U[n] - U_p[n]$, where $U_e[n]$ denotes the error, and $U_p[n]$ is predicted $\lambda_d$ from the prior time slot. In this situation, the coefficients modify through this rule: $h_{n+1}[k] = h_n[k] + \mu U_e[n] \, U[n-k]$ where $\mu$ is the step size that is a critical parameter since it tunes the convergence speed of the algorithm. Generally, adaptive filters (such as LMS) strive to learn from the previous values and because of this they keep the history, so there is no reason for concern about individual traces.

**Remark 2.** With reference to Assumption 3-1, we can estimate average arrival rate $\lambda_t$ simply by an estimation algorithm (see Algorithm 1). When a coding node detects a MAC layer transmission opportunity, it updates a counter that stores the total number of transmission opportunities and $\lambda_t$ is then updated.

**Algorithm 1:** Online $\lambda_t$ estimation
```
if (transmissionOpportunityOccurred() = true)
{
        transmissionCount++;
        λt = transmissionCount/getTimeElapsed();
}
```

# 5- Numerical Results and Discussion

In this section, we describe how our proposed algorithm is evaluated through simulation. In Section 5.1, we present the simulation setup parameters and in Section 5.2, we investigate the simulation results.

## 5-1- Simulation Environment

Simulation of the proposed scheme, COPE [4] and CORE [core] was run on the NS2 simulator [35] which allowed us to compare our scheme's performance with those of COPE (as the basic approach) and CORE (as an approach with some improvements in throughput). In particular, CORE tries to improve network throughput by combing hop-by-hop opportunistic forwarding and network coding in wireless multi-hop networks. The priority of the candidates in the forwarding set in CORE is established based on the coding patterns. The CORE's system model is similar to the proposed scheme's model (and the COPE's model), it has been developed for multiple unicast flows in multi-hop wireless networks using inter-flow network coding, and hence CORE is one of the best options to employ for comparison purposes.

We considered 200 static nodes that were deployed in an $1100 \times 1100$ m$^2$ square field and we modeled the signal attenuation through the two-ray ground propagation model and wireless links with i.i.d. Rayleigh distribution with the same average SNR. Next, we employed UDP traffic, where the packet was set to 1000 B. We did not generate any TCP traffic at the transport layer, instead, in order to monitor the proposed coding solution efficiency at the MAC layer, we used UDP traffic in our simulation scenarios. The experimental results [4] reveal that when the traffic does not exercise congestion control (e.g. UDP), coding throughput improvement may substantially exceed the expected theoretical coding gain. We changed the network offered load by manipulating the number of flows in the network. Flows were established randomly between two nodes as source and destination. Traffic load was generated at all sources using exponential distribution of inter-arrival times with the same intensity. We also used a very simple geographic routing, so the routing strategy selected the neighbor of the node nearest to the final destination as the next hop.

Also in this set up, we applied the power consumption model in [36], as shown in Table 4. Nodes have a nominal transmission range of $\rho$=200m and transmissions are received by all the nodes within transmission range. In followings, we set $L$=40 packets, delay discount factor $\delta$=0.05 and investigate a linear gain function $g(d)= d$. In fact, when a coded packet is transmitted with coding degree $d$, it is reduced $d$-1 transmissions in the network and as a result we set $b$=0 and $c$=1. As has been argued already, without loss of generality, we used LMS adaptive filter for $\lambda_d$ online estimation (see Remark 1). Based on our simulation results, we found that the LMS adaptive filter outperforms the best results with four taps in our scheme.

To make performance comparison, three competing schemes were examined on a variety of network settings. The most important difference between implementations of our scheme and other two algorithms was that in the proposed scheme each coding node followed the decision rule $w$ at each transmission opportunity. In particular, the coding node compared the best coding option degree value with $d^*$; if the pattern coding degree value was greater than or equal to $d^*$, then the coding node transmitted the packet, otherwise the coding node waited for the next transmission opportunity in hope of better coding options. In contrast, in both COPE and CORE, when a transmission opportunity arose for the coding node, the packets were sent immediately with any coding degree value.

## 5-2- Simulation Results

We present the simulation results in this section. Results are obtained by taking average over 100 runs, each of which ends at 30,000 time units. Four important performance metrics are evaluated which include: (i) Average coding gain, (ii) Average end-to-end delay (iii) Throughput and (iv) Average energy consumption per node.

### 5-2-1- Average Coding Gain

Network coding gain was defined as the ratio of the number of transmissions in the non-coding scheme, to the minimum number of transmissions in the coding scheme that delivered the same set of packets to the next-hops successfully (we didn't consider any failed transmissions). Figure 4 compares proposed scheme coding gain against COPE and CORE under various conditions. As can be seen, by increasing the offered load, coding gain rises all three schemes. Existing smooth breakages in proposed scheme diagrams are due to increases in $d^*$. For example, when there are transitions of $d^*$ from the values lower than 1 to those above 1, proposed scheme does not transmit the packets until it has reached coding degree 2. For this reason, average coding gain some times exhibits a step-wise behaviour in response to increases in offered traffic load. In addition, it is seen that CORE outperforms COPE because it provides more coding opportunities via opportunistic forwarding.

### 5-2-2- End-to-End Average Delay

We defined end-to-end as the sum of all delays caused by the network operation during packet transmission, such as buffering latency, queuing operations latency, MAC layer operations latency, routing operation latency, and propagation latency. In practice, if the number of flows across the whole network is small, the coding opportunities are few, so both coding schemes and non-coding scheme operate likewise. But, as the number of flows climbs, the coding opportunities increases too, and it is then that we cannot ignore the collisions in congested wireless networks. Figure 5 shows that the end-to-end average delay increases with increases in offered load in investigated schemes, which is due to the increase in collision in MAC layer. However, average end-to-end delay in proposed scheme is longer than that in COPE, especially when $d^*$ increases by one unit.

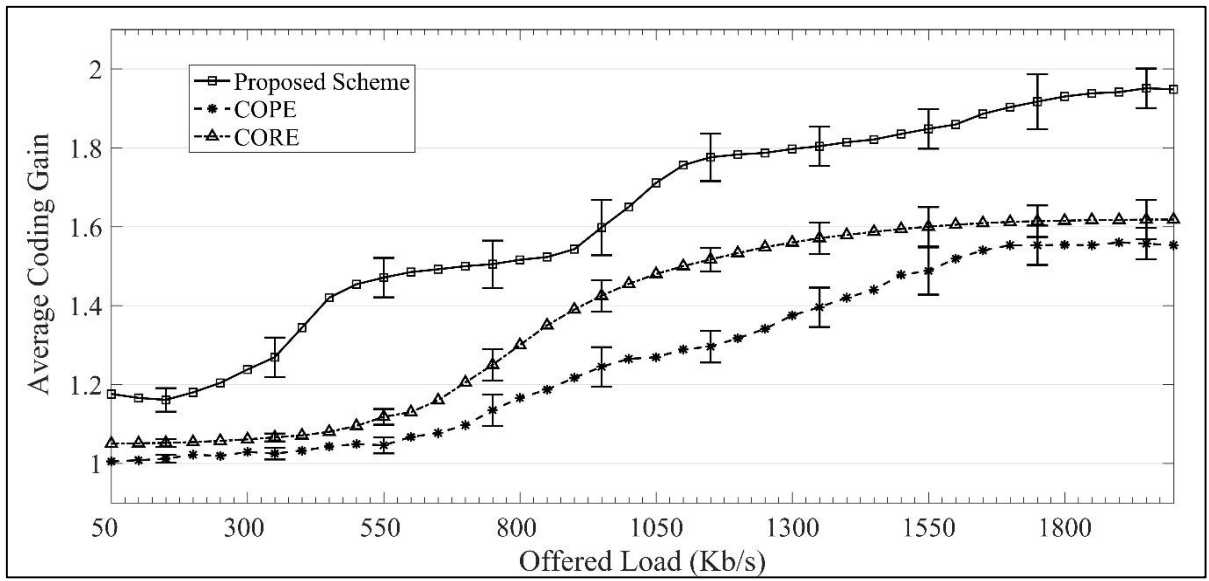

Figure 4: Average coding gain comparison with varying offered load

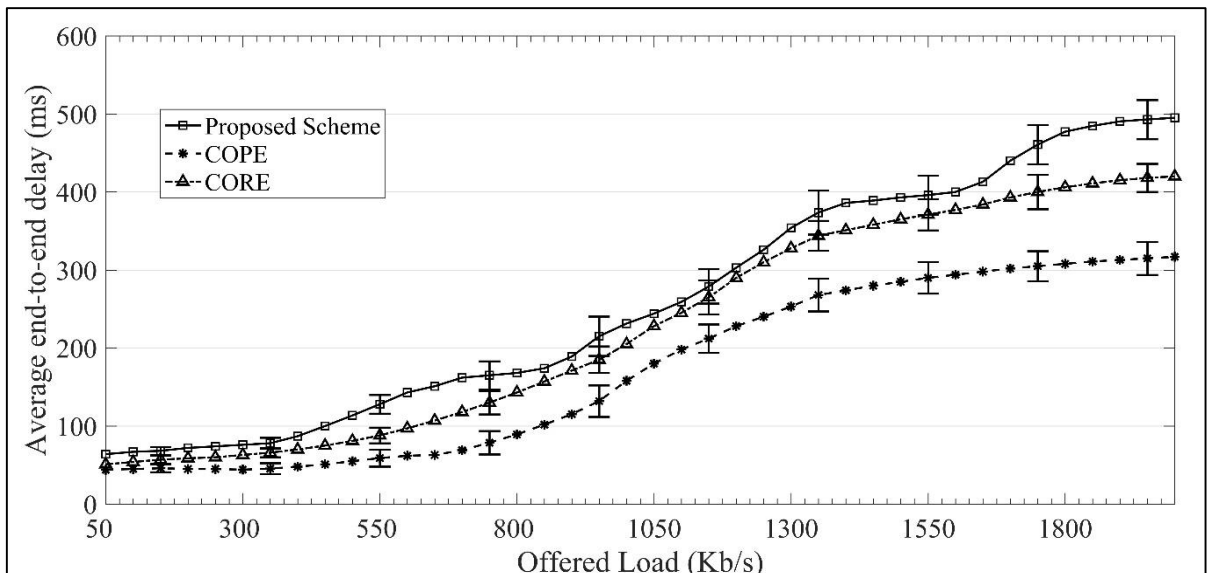

Figure 5: Average end-to-end delay under various offered loads

### 5-2-3- Throughput

We defined throughput as the average rate at which packets are delivered across receivers. In Figure 6, throughput of the proposed scheme is compared against those of COPE and CORE. When the offered load is small, the probability that flows cross each other is small. In contrast, as the number of flows grows, much more coding opportunities are created. The throughput then transitions into a saturated region where the allocated load for each node exceeds a threshold. With the offered load increasing further, the network throughput decreases slowly because of congestion and thus more link-layer retransmissions. In proposed scheme, by increasing offered load and due to increased average coding degree, the number of transmissions decreases for specific volume of traffic, which generally leads to increased throughput of the network. When the number of flows continues to grow, congestion and packet collision also embitter all protocols.

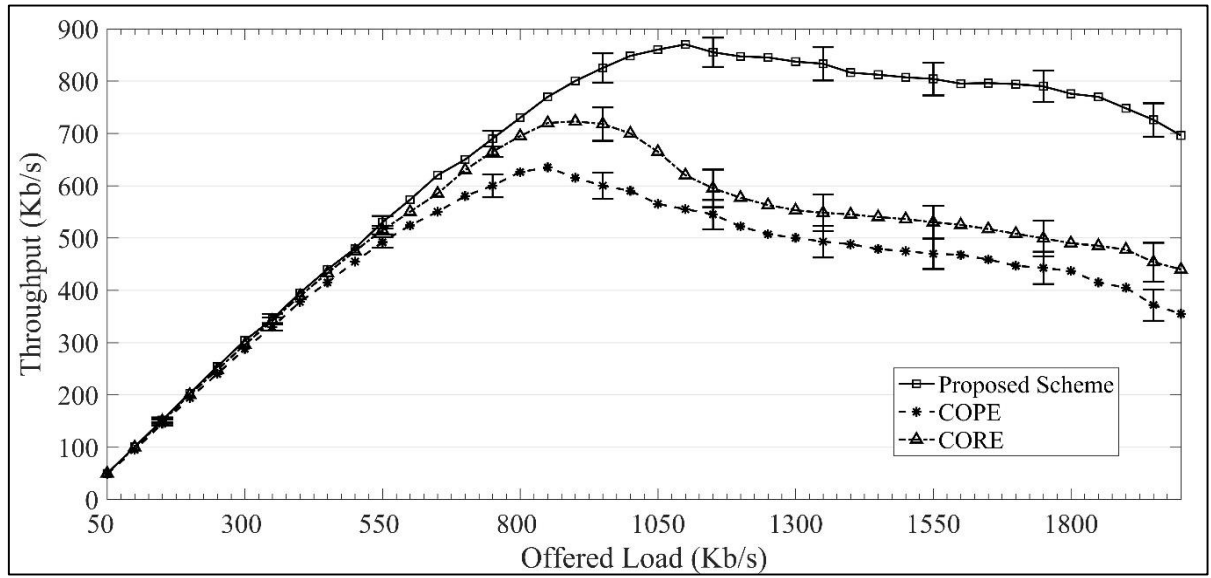

**Figure 6: Throughput: COPE vs. proposed scheme**

**5-2-4- Energy Consumption**

The average energy consumption per node was obtained by dividing the entire network energy consumption by the total number of nodes. The network nodes consume energy for all their activities including sending and receiving packets (whether codded packets or pure packets). By increasing offered load, the average energy consumed by nodes increased. Considering various studies (such as [36]), the amount of energy consumed in radio unit for transmission is always more than the amount for receiving, and the consumption for receiving far exceeds the amount the network used in idle mode. Network coding reduces the energy consumption in wireless networks through combining data packets together and reducing the number of transmission. What is important to note is that although the total number of transmissions in proposed scheme is far lower than that in COPE, energy is still consumed to power up electronic circuits (such as processor and memory) in idle periods. Therefore, the ratio of energy consumed in transmitting mode to that consumed in idle mode has a crucial effect on saving energy. In general, since consumption in transmission mode outweighs that in idle mode, proposed scheme (which delays transmissions until the number of transmissions decreases) is more energy efficient. Also, CORE increases throughput against COPE, therefore CORE saves more energy. Figure 7 illustrates average consumption per node with the assumptions in Table 4. The results includes the amount of energy consumed by both the coded packets and the pure packets transmitted through medium.

**Table 4: The power consumption model**

| | |
|---|---|
| Radio transmitter | 70 mW |
| Radio receiver | 50 mW |
| Radio idle | 25 mW |
| Circuit components | 10 mW |

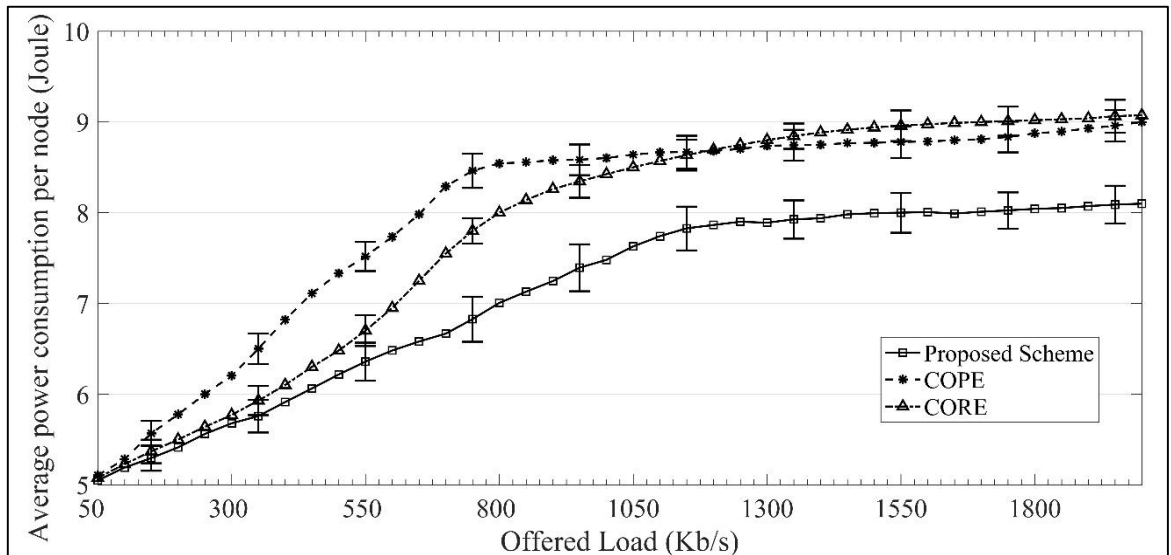

Figure 7: Average energy consumption per node

### 5-2-5- Effect of Channel Error on Performance

In this section we study the impact of channel error on performance of the two approaches, COPE and proposed scheme. In the previous simulation scenarios, we set SNR to a high value (about 25db), thus BER (bit error rate) was negligible. It was observed that the higher the SNR, the lower the BER. A decrease in BER will result in less packet losses at the network, consequently the network throughput of each scheme increases as SNR increases. However, at times of high channel error, and due to more packet loss, reception reports may arrive at the receiver too late totally. Under these circumstances, coded packets are transmitted by a lower coding degree average. We assume that all links have the same average output SNR and an independent Rayleigh fading for each link. In the current simulation scenario, we vary the average SNR at destination nodes in order to assess the impact of channel error on the performance parameters. Figure 8 displays throughput gain of the two schemes with the received SNR for two different traffic load. As can be seen, a higher throughput is achieved for high SNR values. In particular, coding gain increases with increases in the received SNR in the proposed scheme and COPE, which is due to decrease in BER and packet loss rate.

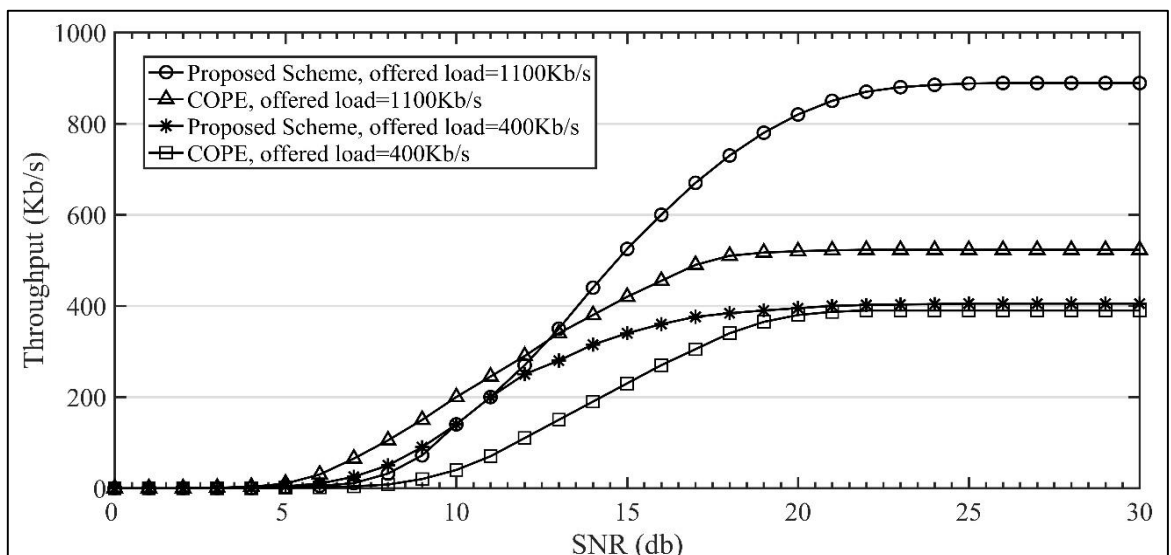

Figure 8: Throughput versus the received SNR

# 6- Conclusion

We have seen that much attention has been given to the study of delayed transmission strategies in two-way relay wireless networks, e.g., in [9-14]. Yet, due to the major challenges in modeling the complex relation between different sessions in multi-hop wireless networks, clearly further research is required with respect to modeling of the to-send-or-not-to-send problem in COPE-based network coding, especially when the traffic flows and the transmission opportunities are stochastic. In this paper, we succeeded in resolving to-sent-or-not-to-send problem based on COPE architecture, which is more general than reverse carpooling scenario. In the proposed algorithm, coding nodes decide whether to postpone transmission of packets in the hope for better coding opportunities to come up (thereby increasing coding degree) or to transmit packets immediately during transmission opportunities in order to decrease end-to-end delay.

We formulated this stochastic sequential decision-making problem as an optimal stopping problem and derived a stopping rule to strike the desired trade-off between number of transmissions and end-to-end delay. In particular, we proved that the 1-SLA rule, which is a simple threshold-based rule, is optimal due to the traffic features. The analytical work then underwent extensive simulations in order to verify its practical feasibility. Our simulation results confirmed that the innovative solution offers significant performance benefits. These benefits are as result of increase in the total throughput and decrease in the total energy consumption. In contrast, the proposed scheme increases the end-to-end average delay.